\documentclass[english]{article}
\usepackage[T1]{fontenc}
\usepackage[latin9]{inputenc}
\usepackage{geometry}
\usepackage{amsmath}
\usepackage{amssymb}
\usepackage{esint}

\makeatletter

\providecommand{\LyX}{L\kern-.1667em\lower.25em\hbox{Y}\kern-.125emX\@}

\@ifundefined{definecolor}
 {\usepackage{color}}{}
\@ifundefined{definecolor}{\usepackage{color}}{}

\usepackage{babel}
\usepackage{nicefrac}
\usepackage{sagetex}

\makeatother

\usepackage{babel}
\begin{document}

\title{A $\textsf{SageTeX}$ Hypermatrix Algebra Package}

\author{Edinah K. Gnang%
\thanks{School of Mathematics, Institute for Advanced Study%
}, Ori Parzanchevski%
\thanks{School of Mathematics, Institute for Advanced Study%
}, Yuval Filmus%
\thanks{School of Mathematics, Institute for Advanced Study%
}}
\maketitle
\begin{abstract}
We describe here a rudimentary sage \cite{S6} implementation of the
Bhattacharya-Mesner hypermatrix algebra package.
\end{abstract}

\section{Introduction}

The current package implements very basic features of the Bhattacharya-Mesner
\emph{hypermatri}x algebra. A hypermatrix denotes a finite set of
complex numbers each of which is indexed by members of an integer
cartesian product set of the form $\left\{ 0,\cdots,\left(n_{0}-1\right)\right\} \times\cdots\times\left\{ 0,\cdots,\left(n_{l-1}-1\right)\right\} $.
Such a hypermatrix is said to be of order $l$ or simply an $l$-hypermatrix
for short. The algebra and the spectral analysis of hypermatrices
arise as a natural generalization of matrix algebra. Important hypermatrix
results available in the literature are concisely surveyed in \cite{L},
the reader is also refered to \cite{LQ} for a more detail survey
on the spectral analysis of hypermatrices. The hypermatrix algebra
discussed here differs from the hypermatrix algebras surveyed in \cite{L}
in the fact that the hypermatrix algebra considered here centers around
the Bhattacharya-Mesner hypermatrix product operation introduced in
\cite{BM1,BM2,B} and followed up in \cite{GER}. Although the scope
of the Bhattacharya-Mesner algebra extends to hypermatrices of all
finite integral orders, the package will be mostly geared towards
$3$-hypermatrices.

\section{The Hypermatrix Sage Package}

We try here to simultaneously follow precepts of the New Jersey school
of experimental mathematics initiated by Doron Zeilberger \cite{Z}
and the fundamental paradigm of litterate programming pioneered by
Donald Knuth\cite{K} to discuss various computational aspects of
the Bhattacharya-Mesner 3-hypermatrix algebra. We therefore present
here a very rudimentary $\textsf{SageTeX}$\cite{S6} implementation
of a hypermatrix package. The proposed package is available through
the source code for the current document either in the format of a
\LyX{} file or alternatively as a $\TeX$ file or an independent sage
file.\\
Our implementation will be concerned with generic 3-hypermatrices
and consequently we will often work with symbolic expressions. The
implementation starts out by describing procedures which enable us
to generate symbolic matrices and hypermatrices of desired size, order
and with other additional properties. Throughout the package, the
data structure used will be a lists.\\
\begin{sageblock}
def MatrixGenerate(nr, nc, c):
    """
    Generates a list of lists associated with a symbolic nr x nc
    matrix using the input character c followed by indices.
    
    EXAMPLES:
    ::
        sage: M = MatrixGenerate(2, 2, 'm'); M
        [[m00, m01], [m10, m11]]

    AUTHORS:
    - Edinah K. Gnang and Ori Parzanchevski
    """
    # Setting the dimensions parameters.
    n_q_rows = nr
    n_q_cols = nc

    # Test for dimension match
    if n_q_rows > 0 and n_q_cols > 0:
        # Initialization of the hypermatrix
        q = []
        for i in range(n_q_rows):
            q.append([])
        for i in range(len(q)):
            for j in range(n_q_cols):
                # Filling up the matrix
                (q[i]).append(var(c+str(i)+str(j)))
        return q

    else :
        raise ValueError, "Input dimensions "+\
str(nr)+" and "+str(nc)+" must both be non-zero positive integers."
\end{sageblock}\\
in addition we implement a similar procedure for generating symbolic
symmetric matrices\\
\begin{sageblock}
def SymMatrixGenerate(nr, c):
    """
    Generates a list of lists associated with a symbolic nr x nc
    symmetric matrix using the input character c followed by 
    indices.
    
    EXAMPLES:
    ::
        sage: M = SymMatrixGenerate(2, 'm'); M
        [[m00, m01], [m10, m11]]

    AUTHORS:
    - Edinah K. Gnang and Ori Parzanchevski
    """
    # Setting the dimensions parameters.
    n_q_rows = nr
    n_q_cols = nr

    # Test for dimension match
    if n_q_rows > 0 and n_q_cols > 0:
        # Initialization of the hypermatrix
        q = []
        for i in range(n_q_rows):
            q.append([])
        for i in range(len(q)):
            for j in range(n_q_cols):
                # Filling up the matrix
                (q[i]).append(var(c+str(min(i,j))+str(max(i,j))))
        return q

    else :
        raise ValueError, "Input dimensions "+\
str(nr)+" must be a non-zero positive integers."
\end{sageblock}\\
The two procedures implemented above for generating symbolic lists
will typically be used in conjunction with the Sage\cite{S6} Matrix
class over symbolic rings as illustrated 
\begin{equation}
\mathbf{M}_{1}=\mbox{Matrix(SR,MatrixGenerate(2,3,'m'))}=\sage{Matrix(SR,MatrixGenerate(2,3,'m'))}.
\end{equation}
\begin{equation}
\mathbf{M}_{2}=\mbox{Matrix(SR,SymMatrixGenerate(2,'m'))}=\sage{Matrix(SR,SymMatrixGenerate(2,'m'))}.
\end{equation}
We implement similar procedures for generating symbolic hypermatrices
of desired order and size.\\
\begin{sageblock}
def HypermatrixGenerate(*args):
    """
    Generates a list of lists associated with a symbolic arbitrary
    hypematrix of order and size specified by the input.

    EXAMPLES:
    ::
        sage: M = HypermatrixGenerate(2, 2, 2, 'm'); M
    
     AUTHORS:
    - Edinah K. Gnang, Ori Parzanchevski and Yuval Filmus
    """
    if len(args) == 1:
        return var(args[0])
    return [apply(\
HypermatrixGenerate,args[1:-1]+(args[-1]+str(i),)) for i in range(args[0])]
\end{sageblock}\\
The procedures implemented above illustrate the use of lists for representing
hypermatrices. We show bellow for convenience of the reader the output
of the function call 
\begin{equation}
\mathbf{T}=\mbox{HypermatrixGenerate(2, 2, 2, 't')}=\sage{HypermatrixGenerate(2,2,2,'t')}.
\end{equation}
In connection with the spectral decomposition of 3-hypermatrices we
discuss the implemention of a procedure which generates the desired
size symbolic 3-hypermatrices with entries symmetric under cyclic
permutation of the hypermatrix indices.\\
\begin{sageblock}
def SymHypermatrixGenerate(nr, c):
    """
    Generates a list of lists associated with a symbolic nr x nc x nd
    third order hypematrix using the input character c followed by 
    indices.
    
    EXAMPLES:
    ::
        sage: M = SymHypermatrixGenerate(2, 'm'); M
        
    AUTHORS:
    - Edinah K. Gnang and Ori Parzanchevski
    """
    # Setting the dimensions parameters.
    n_q_rows = nr
    n_q_cols = nr
    n_q_dpts = nr

    # Test for dimension match
    if n_q_rows > 0 and n_q_cols > 0 and n_q_dpts >0:
        # Initialization of the hypermatrix
        q = []
        for i in range(n_q_rows):
            q.append([])
        for i in range(len(q)):
            for j in range(n_q_cols):
                (q[i]).append([])
        for i in range(len(q)):
            for j in range(len(q[i])):
                for k in range(n_q_dpts):
                    if i==j or i==k or j==k:
                        (q[i][j]).append(\
var(c+str(min(i,j,k))+str(i+j+k-min(i,j,k)-max(i,j,k))+str(max(i,j,k))))
                    else:
                        if i == min(i,j,k) and k == max(i,j,k):
                            (q[i][j]).append(\
var(c+str(min(i,j,k))+str(i+j+k-min(i,j,k)-max(i,j,k))+str(max(i,j,k))))
                        elif k == min(i,j,k) and j == max(i,j,k):
                            (q[i][j]).append(\
var(c+str(min(i,j,k))+str(i+j+k-min(i,j,k)-max(i,j,k))+str(max(i,j,k))))
                        elif i == max(i,j,k) and j == min(i,j,k):
                            (q[i][j]).append(\
var(c+str(min(i,j,k))+str(i+j+k-min(i,j,k)-max(i,j,k))+str(max(i,j,k))))
                        else:
                            (q[i][j]).append(\
var(c+str(i+j+k-min(i,j,k)-max(i,j,k))+str(min(i,j,k))+str(max(i,j,k))))
        return q

    else :
        raise ValueError, "Input dimensions "+\
str(nr)+" must be a non-zero positive integer."
\end{sageblock}\\
We illustrate the use of the procedure by showing the output of the
following function call 
\begin{equation}
\mathbf{S}=\mbox{SymHypermatrixGenerate(2, 's')}=\sage{SymHypermatrixGenerate(2,'s')}.
\end{equation}
We also implement a procedure for canonically stripping down the 3-hypermatrix
( encoded as a list of list ) to a simple list of symbolic variables
in a similar spirit as the matrix vectorization operation.\\
\begin{sageblock}
def HypermatrixVectorize(A):
    """
    Outputs our canonical vectorization of
    the input hypermatrices A.
    
    EXAMPLES:
    ::
        sage: M = HypermatrixVectorize(A); M
        

    AUTHORS:
    - Edinah K. Gnang and Ori Parzanchevski
    """
    # Setting the dimensions parameters.
    n_q_rows = len(A)
    n_q_cols = len(A[0])
    n_q_dpts = len(A[0][0])

    # Test for dimension match
    if n_q_rows>0 and n_q_cols>0 and n_q_dpts>0:
        # Initialization of the hypermatrix
        q = []
        for i in range(n_q_rows):
            for j in range(n_q_cols):
                for k in range(n_q_dpts):
                    q.append(A[i][j][k])
        return q

    else :
        raise ValueError, "The Dimensions non zero."
\end{sageblock} \\
The implementation of the hypermatrix vectorization procedure concludes
the implementation of procedure for generating and formating symbolic
3-hypermatrices.

The next part of the package will discuss the implementation of procedures
which enable us to perform very basic operations on 3-hypermatrices
starting with the addition operation\\
\begin{sageblock}
def HypermatrixAdd(A, B):
    """
    Outputs a list of lists corresponding to the sum of 
    the two input hypermatrices A, B of the same size
    
    EXAMPLES:
    ::
        sage: M = HypermatrixAdd(A, B); M
        

    AUTHORS:
    - Edinah K. Gnang and Ori Parzanchevski
    """
    # Setting the dimensions parameters.
    n_q_rows = len(B)
    n_q_cols = len(B[0])
    n_q_dpts = len(B[0][0])

    # Test for dimension match
    if n_q_rows==len(A) and n_q_cols==len(A[0]) and n_q_dpts==len(A[0][0]):
        # Initialization of the hypermatrix
        q = []
        for i in range(n_q_rows):
            q.append([])
        for i in range(len(q)):
            for j in range(n_q_cols):
                (q[i]).append([])
        for i in range(len(q)):
            for j in range(len(q[i])):
                for k in range(n_q_dpts):
                    (q[i][j]).append(A[i][j][k]+B[i][j][k])
        return q

    else :
        raise ValueError, "The Dimensions of the input hypermatrices must match."
\end{sageblock}\\
quite similarly we implement the 3-hypermatrix hadamard product procedure\\
\begin{sageblock}
def HypermatrixHadamardProduct(A, B):
    """
    Outputs a list of lists associated with the addtion of 
    the two input hypermatrices A and B
    
    EXAMPLES:
    ::
        sage: M = HypermatrixHadamardProduct(A, B); M
        

    AUTHORS:
    - Edinah K. Gnang and Ori Parzanchevski
    """
    # Setting the dimensions parameters.
    n_q_rows = len(A)
    n_q_cols = len(A[0])
    n_q_dpts = len(A[0][0])

    # Test for dimension match
    if n_q_rows==len(A) and n_q_cols==len(A[0]) and n_q_dpts==len(A[0][0]):
        # Initialization of the hypermatrix
        q = []
        for i in range(n_q_rows):
            q.append([])
        for i in range(len(q)):
            for j in range(n_q_cols):
                (q[i]).append([])
        for i in range(len(q)):
            for j in range(len(q[i])):
                for k in range(n_q_dpts):
                    (q[i][j]).append(A[i][j][k]*B[i][j][k])
        return q

    else :
        raise ValueError, "The Dimensions of the input hypermatrices must match."
\end{sageblock}\\
We illustrate the usage of the two procedures implemented above 
\[
\mathbf{S}+\mathbf{T}=\mbox{HypermatrixAdd(SymHypermatrixGenerate(2,'s'),HypermatrixGenerate(2,2,2,'t'))}=
\]
\begin{equation}
\sage{HypermatrixAdd(SymHypermatrixGenerate(2,'s'),HypermatrixGenerate(2,2,2,'t'))}
\end{equation}
and
\[
\mathbf{S}\star\mathbf{T}=\mbox{HypermatrixHadamardProduct(SymHypermatrixGenerate(2,'s'),HypermatrixGenerate(2,2,2,'t'))}=
\]
\begin{equation}
\sage{HypermatrixHadamardProduct(SymHypermatrixGenerate(2,'s'),HypermatrixGenerate(2,2,2,'t'))}.
\end{equation}
Furthermore, we implement the procedure for multiplying a 3-hypermatrix
by a scalar.\\
\begin{sageblock}
def HypermatrixScale(A, s):
    """
    Outputs a list of lists associated with product of the
    input scalar s with the input hypermatrix A.
    
    EXAMPLES:
    ::
        sage: M = HypermatrixScale(A, 3); M
        

    AUTHORS:
    - Edinah K. Gnang and Ori Parzanchevski
    """
    # Setting the dimensions parameters.
    n_q_rows = len(A)
    n_q_cols = len(A[0])
    n_q_dpts = len(A[0][0])

    # Initialization of the hypermatrix
    q = []
    for i in range(n_q_rows):
        q.append([])
    for i in range(len(q)):
        for j in range(n_q_cols):
            (q[i]).append([])
    for i in range(len(q)):
        for j in range(len(q[i])):
            for k in range(n_q_dpts):
                (q[i][j]).append(A[i][j][k]*s)
    return q
\end{sageblock}\\
typically used as follows 
\[
3\,\mathbf{T}=\mbox{HypermatrixScale(HypermatrixGenerate(2,2,2,'t'),3)}=
\]
\begin{equation}
\sage{HypermatrixScale(HypermatrixGenerate(2,2,2,'t'),3)}.
\end{equation}
similarly, we implement the entry-wise exponentiation bellow\\
\begin{sageblock}
def HypermatrixEntryExponent(A, s):
    """
    Outputs a list of lists associated with product of the
    scalar s with the hypermatrix A.
    
    EXAMPLES:
    ::
        sage: M = HypermatrixEntryExponent(A, 3); M
        

    AUTHORS:
    - Edinah K. Gnang and Ori Parzanchevski
    """
    # Setting the dimensions parameters.
    n_q_rows = len(A)
    n_q_cols = len(A[0])
    n_q_dpts = len(A[0][0])

    # Initialization of the hypermatrix
    q = []
    for i in range(n_q_rows):
        q.append([])
    for i in range(len(q)):
        for j in range(n_q_cols):
            (q[i]).append([])
    for i in range(len(q)):
        for j in range(len(q[i])):
            for k in range(n_q_dpts):
                (q[i][j]).append((A[i][j][k])^s)
    return q
\end{sageblock}\\
due to the fact that the exponentiation operation is noncommutative
we also implement the entry-wise exponentiation operation where the
input is to be taken as basis for the exponentiation computation.\\
\begin{sageblock}
def HypermatrixEntryExponentB(s, A):
    """
    Outputs a list of lists associated with product of the
    scalar s with the hypermatrix A.
    
    EXAMPLES:
    ::
        sage: M = HypermatrixEntryExponentB(3,A); M
        
    AUTHORS:
    - Edinah K. Gnang and Ori Parzanchevski
    """
    # Setting the dimensions parameters.
    n_q_rows = len(A)
    n_q_cols = len(A[0])
    n_q_dpts = len(A[0][0])

    # Initialization of the hypermatrix
    q = []
    for i in range(n_q_rows):
        q.append([])
    for i in range(len(q)):
        for j in range(n_q_cols):
            (q[i]).append([])
    for i in range(len(q)):
        for j in range(len(q[i])):
            for k in range(n_q_dpts):
                (q[i][j]).append(s^(A[i][j][k]))
    return q
\end{sageblock}\\
At the heart of the Mesner-Bhattacharya 3-hypermatrix algebra lies
the ternary non-associative hypermatrix product operation\cite{BM2,BM1}.
We provide here a naive implementation of the Mesner-Bhattacharya
3-hypermatrix product. We may briefly recall that the product is defined
for input hypermatrices $\mathbf{A}$ of dimensions $m\times k\times p$,
$\mathbf{B}$ of dimensions $m\times n\times k$ and the matrix $\mathbf{C}$
of dimension $k\times n\times p$, to result into an $m\times n\times p$
hypermatrix with entries specified by 
\begin{equation}
\left[\circ\left(\mathbf{A},\,\mathbf{B},\,\mathbf{C}\right)\right]_{i,j,k}=\sum_{0\le t<k}a_{itk}\, b_{ijt}\, c_{tjk}
\end{equation}
\begin{sageblock}
def HypermatrixProduct(A, B, C):
    """
    Outputs a list of lists associated with the ternary 
    non associative Bhattacharya-Mesner product of the
    input hypermatrices A, B and C.
    
    EXAMPLES:
    ::
        sage: M = HypermatrixProduct(A, B, C); M
        
    AUTHORS:
    - Edinah K. Gnang and Ori Parzanchevski
    """
    # Setting the dimensions parameters.
    n_a_rows = len(A)
    n_a_cols = len(A[0])
    n_a_dpts = len(A[0][0])

    n_b_rows = len(B)
    n_b_cols = len(B[0])
    n_b_dpts = len(B[0][0])

    n_c_rows = len(C)
    n_c_cols = len(C[0])
    n_c_dpts = len(C[0][0])

    # Test for dimension match
    if n_a_rows==n_b_rows and n_b_cols==n_c_cols and n_c_dpts==n_a_dpts and \
n_a_cols==n_b_dpts and n_b_dpts==n_c_rows:
        # Initialization of the hypermatrix
        q = []
        for i in range(n_a_rows):
            q.append([])
        for i in range(len(q)):
            for j in range(n_b_cols):
                (q[i]).append([])
        for i in range(len(q)):
            for j in range(len(q[i])):
                for k in range(n_c_dpts):
                    (q[i][j]).append(\
sum([A[i][l][k]*B[i][j][l]*C[l][j][k] for l in range(n_a_cols)]))
        return q

    else :
        raise ValueError, "Hypermatrix dimension mismatch."
\end{sageblock}\\
In connection with the computation of the spectral elimination ideals,
we implement a slight generalization of the Mesner-Bhattacharya hypermatrix
product hypermatrix product, introduced in \cite{GER}. Recall that
the 3-hypermatrix product of input hypermatrices $\mathbf{A}$ of
dimensions $m\times l\times p$, $\mathbf{B}$ of dimensions $m\times n\times l$
and the matrix $\mathbf{C}$ of dimension $l\times n\times p$, with
non-trivial background $\mathbf{T}$ with dimensions $l\times l\times l$
results in $m\times n\times p$ hypermatrix and in particular the
$m$, $n$, $p$ of the product is expressed by 
\begin{equation}
\left[\circ_{\mathbf{T}}\left(\mathbf{A},\,\mathbf{B},\,\mathbf{C}\right)\right]_{mnp}=\sum_{1\le i\le l}\left(\sum_{1\le j\le l}\left(\sum_{1\le k\le l}a_{mip}\, b_{mnj}\, c_{knp}\, t_{ijk}\right)\right),
\end{equation}
which is implemented as follows\\
\begin{sageblock}
def HypermatrixProductB(A, B, C, D):
    """
    Outputs a list of lists associated with the ternary 
    product the input hypermatrices A, B and C with
    background hypermatrix D.
    
    EXAMPLES:
    ::
        sage: M = HypermatrixProductB(A, B, C, D); M
        
    AUTHORS:
    - Edinah K. Gnang and Ori Parzanchevski
    """
    # Setting the dimensions parameters.
    n_a_rows = len(A)
    n_a_cols = len(A[0])
    n_a_dpts = len(A[0][0])

    n_b_rows = len(B)
    n_b_cols = len(B[0])
    n_b_dpts = len(B[0][0])

    n_c_rows = len(C)
    n_c_cols = len(C[0])
    n_c_dpts = len(C[0][0])

    n_d_rows = len(D)
    n_d_cols = len(D[0])
    n_d_dpts = len(D[0][0])

    # Test for dimension match
    if \
n_a_rows==n_b_rows and n_b_cols==n_c_cols and n_c_dpts==n_a_dpts and \
n_a_cols==n_b_dpts and n_b_dpts==n_c_rows and n_a_cols==n_d_rows and \
n_a_cols==n_d_cols and n_a_cols==n_d_dpts:
        # Initialization of the hypermatrix
        q = []
        for i in range(n_a_rows):
            q.append([])
        for i in range(len(q)):
            for j in range(n_b_cols):
                (q[i]).append([])
        for i in range(len(q)):
            for j in range(len(q[i])):
                for k in range(n_c_dpts):
                    (q[i][j]).append(\
sum([A[i][l0][k]*B[i][j][l1]*C[l2][j][k]*D[l0][l1][l2] for l0 in range(n_d_rows)\
for l1 in range(n_d_cols) for l2 in range(n_d_dpts)]))
        return q

    else :
        raise ValueError, "Hypermatrix dimension mismatch."
\end{sageblock}\\
We illustrate bellow, how to initialize and obtain 3-hypermatrix products
either with the trivial or arbitrary background hypermatrix. The example
discussed here will be for $2\times2\times2$ hypermatrices.\\
\begin{sageblock}
# We put here together the seperate pieces we have implemented above.
A = HypermatrixGenerate(2, 2, 2, 'a')
B = HypermatrixGenerate(2, 2, 2, 'b')
C = HypermatrixGenerate(2, 2, 2, 'c')
T = HypermatrixGenerate(2, 2, 2, 't')
P = HypermatrixProduct(A, B, C)
Q = HypermatrixProductB(A, B, C, T)
\end{sageblock}\\
from which we obtain that the $0$,$0$,$0$ entry of the product
with trivial background is given by 
\begin{equation}
p_{000}=\left[\circ\left(\mathbf{A},\,\mathbf{B},\,\mathbf{C}\right)\right]_{0,0,0}=\mbox{P[0][0][0]}=\sage{P[0][0][0]}
\end{equation}
while the $0$,$0$,$0$ entry of the product with non trivial background
is given 
\[
q_{000}=\left[\circ_{\mathbf{T}}\left(\mathbf{A},\,\mathbf{B},\,\mathbf{C}\right)\right]_{0,0,0}=\mbox{Q[0][0][0]}=\sage{sum(((Q[0][0][0]).operands())[0:4])}+
\]
\begin{equation}
\sage{sum(((Q[0][0][0]).operands())[4:8])}.
\end{equation}
We now implement the procedure which generalizes to 3-hypermatrices
the notion of matrix transpose. The transpose operation for matrices
consists in performing a transposition of matrix indices and this
has the effect of simultaneously changing rows vectors into column
vectors and column vecors into row vectors. However in the case of
3-hypermatrices there are six possible permutations which can be performed
on the indices and among these permutations, the cyclic permutation
form a very special subgroup, because cyclic permutations simultaneously
map rows vectors to columns vectors and column vectors to depth vectors.
As a result, cyclic permutations of the indices should be thought
off as operations which are inherent to 3-hypermatrices while the
remaining three transpositions are to be thought off as matrix operations.\\
\begin{sageblock}
def HypermatrixCyclicPermute(A):
    """
    Outputs a list of lists associated with the hypermatrix 
    with entries index cycliclly permuted.
    
    EXAMPLES:
    ::
        sage: M = HypermatrixCyclicPermute(A); M
        

    AUTHORS:
    - Edinah K. Gnang and Ori Parzanchevski
    """
    # Setting the dimensions parameters.
    n_q_rows = len(A[0])
    n_q_cols = len(A[0][0])
    n_q_dpts = len(A)

    # Initialization of the hypermatrix
    q = []
    for i in range(n_q_rows):
        q.append([])
    for i in range(len(q)):
        for j in range(n_q_cols):
            (q[i]).append([])
    for i in range(len(q)):
        for j in range(len(q[i])):
            for k in range(n_q_dpts):
                (q[i][j]).append(A[k][i][j])
    return q
\end{sageblock}\\
We illustrate the hypermatrix transpose operation by starting with
the 3-hypermatrix 
\begin{equation}
\mathbf{A}=\sage{A}
\end{equation}
and showing the result of the transposition 
\begin{equation}
\mathbf{A}^{T}=\mbox{HypermatrixCyclicPermute(A)}=\sage{HypermatrixCyclicPermute(A)}.
\end{equation}
In connection with 3-hypermatrix spectral decompositions computations,
we implement procedure for generating special family of 3-hypermatrices
starting with Kronecker delta 3-hypermatrices. The defining properties
of the Kronecker delta 3-hypermatrix can be expressed as follows 
\begin{equation}
\boldsymbol{\Delta}=\left(\delta_{ijk}\ge0\right)_{0\le i,j,k<n},\quad\mbox{ and }\quad\boldsymbol{\Delta}=\circ\left(\boldsymbol{\Delta},\,\boldsymbol{\Delta}^{T^{2}},\,\boldsymbol{\Delta}^{T}\right)
\end{equation}
and the procedure generating Kronecker delta 3-hypermatrices is implemented
as follows\\
\begin{sageblock}
def HypermatrixKroneckerDelta(nr):
    """
    Generates a list of lists associated with the nr x nr x nr
    Kronecker Delta hypermatrix.
    
    EXAMPLES:
    ::
        sage: M = HypermatrixKroneckerDelta(2); M
        

    AUTHORS:
    - Edinah K. Gnang and Ori Parzanchevski
    """
    # Setting the dimensions parameters.
    n_q_rows = nr
    n_q_cols = nr
    n_q_dpts = nr

    # Test for dimension match
    if n_q_rows > 0 and n_q_cols > 0 and n_q_dpts >0:
        # Initialization of the hypermatrix
        q = []
        for i in range(n_q_rows):
            q.append([])
        for i in range(len(q)):
            for j in range(n_q_cols):
                (q[i]).append([])
        for i in range(len(q)):
            for j in range(len(q[i])):
                for k in range(n_q_dpts):
                    if i==j and i==k:
                        (q[i][j]).append(1)
                    else:
                        (q[i][j]).append(0)
        return q

    else :
        raise ValueError, "Input dimensions "+\
str(nr)+" must be a non-zero positive integer."
\end{sageblock}\\
Furthermore for some particular numerical routines we implement procedures
for initializing hypermatrices so as to have all entries either equal
to zero or equal to one\\
\begin{sageblock}
def HypermatrixGenerateAllOne(*args):
    """
    Generates a list of lists associated with the nr x nr x nr
    all one hypermatrix.
    
    EXAMPLES:
    ::
        sage: M = HypermatrixGenerateAllOne(2,2,2); M
        

    AUTHORS:
    - Edinah K. Gnang and Ori Parzanchevski
    """
    if len(args) == 1:
        return [1 for i in range(args[0])]
    return [apply(HypermatrixGenerateAllOne, args[1:] ) for i in range(args[0])]
\end{sageblock}\\
for initializing all entries to zero we have\\
\begin{sageblock}
def HypermatrixGenerateAllZero(*args):
    """
    Generates a list of lists associated with the nr x nr x nr
    all zero hypermatrix.
    
    EXAMPLES:
    ::
        sage: M = HypermatrixGenerateAllZero(2,2,2); M
        

    AUTHORS:
    - Edinah K. Gnang and Ori Parzanchevski
    """
    if len(args) == 1:
        return [0 for i in range(args[0])]
    return [apply(HypermatrixGenerateAllZero, args[1:] ) for i in range(args[0])]
\end{sageblock}\\
More interestingly, we implement procedures for generating 3-hypermatrices
with binary entries which correspond to the 3-hypermatrix analogue
of permutation matrices. Permutation 3-hypermatrices by analogy to
permutation matrices effect some prescribed permutations of row slices
or column slices or alternatively the depth slices of some specified
hypermatrices. The permutation is effected by performing the appropriate
sequence hypermatrix products. The procedure which we implement here
for generating permutation 3-hypermatrix takes as input a list of
integer in the range $0$ to $\left(n-1\right)$ inclusively whose
particular order in the list specify the desired transposition. The
procedure outputs the corresponding transposition 3-hypermatrix. The
output 3-hypermatrix will be of dimension $n\times n\times n$. We
recall from \cite{GER} that permutation hypermatrices corresponding
to some transposition $\sigma\in S_{n}$ is expressed by 
\begin{equation}
\mathbf{P}_{\sigma}=\sum_{1\le k\le n}\circ\left(\boldsymbol{1}_{n\times n\times n},\,\boldsymbol{1}_{n\times n\times n},\,\mathbf{e}_{k}\otimes\mathbf{e}_{k}\otimes\mathbf{e}_{\sigma\left(k\right)}\right)
\end{equation}
\begin{sageblock}
def HypermatrixPermutation(s):
    """
    Generates a list of lists associated with the permutation
    hypermatrix deduced from sigma.
    
    EXAMPLES:
    ::
        sage: M = HypermatrixPermutation([0,2,1]); M
        

    AUTHORS:
    - Edinah K. Gnang and Ori Parzanchevski
    """
    n = len(s)
    # Setting the dimensions parameters.
    n_q_rows = n
    n_q_cols = n
    n_q_dpts = n

    # Test for dimension match
    if n_q_rows > 0 and n_q_cols > 0 and n_q_dpts >0:
        # Initialization of the hypermatrix
        q = []
        T = HypermatrixKroneckerDelta(n)
        U = HypermatrixGenerateAllOne(n,n,n)
        Id= HypermatrixProduct(U,U,T)
        Id= HypermatrixCyclicPermute(Id)
        for i in range(n):
            q.append(Id[s[i]])
        return HypermatrixCyclicPermute(HypermatrixCyclicPermute(q))

    else :
        raise ValueError, "Input dimensions "+\
str(n)+" must be a non-zero positive integer."
\end{sageblock}\\
It is important to note that because of the associativity symmetry
breaking, it is important to express the permutations as product of
transpositions. We also illustrate how the 3-hypermatrix product effects
some desired transposition to the appropriate 3-hypermatrix slices.\\
\begin{sageblock}
# the code writen here is merely to put together the peices we have implemented so far.
# Generic Symbolic hypermatrix
A   = HypermatrixGenerateAllZero(3,3,3)  
Tmp = HypermatrixGenerate(3, 3, 3, 'a')
for i in range(2):
    for j in range(3):
        for k in range(3):
            A[i][j][k]=Tmp[i][j][k]

# Initialization of the hypermatrix and it's cyclic permutations
P  = HypermatrixPermutation([1,0,2])
Pt =HypermatrixCyclicPermute(P)
Ptt=HypermatrixCyclicPermute(HypermatrixCyclicPermute(P))

# Effecting the permutation of ...
# row slices
Ar = HypermatrixProduct(Pt,Ptt,A)
# column slice
Ac = HypermatrixProduct(A,P,Pt)
# and depth slices
Ad = HypermatrixProduct(P,A,Ptt)
\end{sageblock} \\
It follows from the lines of code written above that starting from
the $3\times3\times3$ symbolic 3-hypermatrix 
\[
\mathbf{A}=\left[\sage{A[0]},\right.
\]
\begin{equation}
\left.\sage{A[1]}\right]
\end{equation}
and for performing the transposition $\left[1,0,2\right]$, we produced
the permutation hypermatrix 
\begin{equation}
\mathbf{P}_{\left[1,0,2\right]}=\sage{P}.
\end{equation}
In order to effect the transposition to the row slices of $\mathbf{A}$
we compute the product 
\[
\circ\left(\mathbf{P}_{\left[1,0,2\right]}^{T},\mathbf{P}_{\left[1,0,2\right]}^{T^{2}},\mathbf{A}\right)=\left[\sage{Ar[0]},\right.
\]
\begin{equation}
\left.\sage{Ar[1]}\right].
\end{equation}
furthermore in order to effect the transposition to the column slices
of $\mathbf{A}$ we compute the product
\[
\circ\left(\mathbf{A},\mathbf{P}_{\left[1,0,2\right]},\mathbf{P}_{\left[1,0,2\right]}^{T}\right)=\left[\sage{Ac[0]},\right.
\]
\begin{equation}
\left.\sage{Ac[1]}\right].
\end{equation}
finally in order to effect the same transposition to the depth slices
of $\mathbf{A}$ we compute the product
\[
\circ\left(\mathbf{P}_{\left[1,0,2\right]},\mathbf{A},\mathbf{P}_{\left[1,0,2\right]}^{T^{2}}\right)=\left[\sage{Ad[0]},\right.
\]
\begin{equation}
\left.\sage{Ad[1]}\right].
\end{equation}
We now implement a procedure for generating 3-hypermatrix analog of
diagonal martrices. We recall that just as for matrices the diagonal
3-hypermatrices are slight variation of the identity permutation 3-hypermatrix
and their defining equality is expressed by 
\begin{equation}
\mathbf{D}^{\star^{3}}=\circ\left(\mathbf{D}^{T},\,\mathbf{D}^{T^{2}},\,\mathbf{D}\right)
\end{equation}
where $\mathbf{D}^{\star^{3}}$ denotes the Hadamard cube power of
$\mathbf{D}$. The procedure that we implement here for generating
a diagonal 3-hypermatrix, takes as input a symmetric generic $n\times n$
symbolic matrix and outputs a $n\times n\times n$ 3-hypermatrix satisfying
the defining equation\\
\begin{sageblock}
def DiagonalHypermatrix(Mtrx):
    """
    Outputs a diagonal third order hypermatrix
    constructed using the input square matrix
    to enforce the symmetry constraint we will
    only take entry from the lower triangular
    part of the input matrix.

     EXAMPLES:
    ::
        sage: var('a00, a11, a01')
        sage: Mtrx = Matrix(Sr,[[a00,a01],[a01,a11]])
        sage: d = DiagonalHypermatrix(Mtrx)

    AUTHORS:
    - Edinah K. Gnang and Ori Parzanchevski
    """
    # Initialization of the dimensions
    n = min(Mtrx.nrows(),Mtrx.ncols())
    n_d_rows = n
    n_d_cols = n
    n_d_dpts = n

    # Initialization of the identity permutations hypermatrix
    D = HypermatrixPermutation(range(n))

    # Filling up the entries of the hypermatrix.
    for i in range(n_d_rows):
        for j in range(n_d_cols):
            for k in range(n_d_dpts):
                if (D[i][j][k] != 0):
                    D[i][j][k] = Mtrx[min(i,k),max(i,k)]
    return D    
\end{sageblock}\\
We illustrate with the following few lines of codes how to generate
a diagonal $2\times2\times2$ hypermatrices and verify their defining
identity\\
\begin{sageblock}
# Generating a diagonal hypermatrices
Mtrx = Matrix(SR,MatrixGenerate(2, 3,"lambda"))
D  = DiagonalHypermatrix(Mtrx)
Dt = HypermatrixCyclicPermute(D)
Dtt= HypermatrixCyclicPermute(HypermatrixCyclicPermute(D))
Dc = HypermatrixProduct(Dt,Dtt,D)
\end{sageblock}\\
hence 
\begin{equation}
\mathbf{D}=\sage{D}
\end{equation}
and we observe that 
\begin{equation}
\circ\left(\mathbf{D}^{T},\,\mathbf{D}^{T^{2}},\,\mathbf{D}\right)=\mbox{HypermatrixProduct(Dt, Dtt, D)}=\sage{Dc}
\end{equation}
and incidentally has the same entries as the hypermatrix $\mathbf{D}^{\star^{3}}$
\begin{equation}
\mathbf{D}^{\star^{3}}=\mbox{HypermatrixEntryExponent(D, 3)}=\sage{HypermatrixEntryExponent(D,3)}
\end{equation}
We now implement procedures which enables us to constrast $2\times2$
, $2\times2\times2$, and so on type hypermatrices wich are orthogonal
in the sense introduced in \cite{GER}.\\
\begin{sageblock}
def Orthogonal2x2x2Hypermatrix(t):
    """
    Outputs an orthogonal third order hypermatrix
    of size 2 by 2 by 2.

     EXAMPLES:
    ::
        sage: t=var('t')
        sage: Orthogonal2x2x2Hypermatrix(t)

    AUTHORS:
    - Edinah K. Gnang and Ori Parzanchevski
    """
    return [[[cos(t)^(2/3),sin(t)^(2/3)],[sin(t)^(2/3), cos(t)^(2/3)]],\
[[-sin(t)^(2/3),cos(t)^(2/3)],[sin(t)^(2/3),sin(t)^(2/3)]]]    
\end{sageblock}\\
we also present here a parametrization of a subset of $3\times3\times3$
orthogonal hypermatrix bellow\\
\begin{sageblock}
def Orthogonal3x3x3Hypermatrix(t1,t2):
    """
    Outputs an orthogonal third order hypermatrix
    of size 3 by 3 by 3.

     EXAMPLES:
    ::
        sage: t1,t2=var('t1,t2')
        sage: Orthogonal3x3x3Hypermatrix(t1,t2)

    AUTHORS:
    - Edinah K. Gnang and Ori Parzanchevski
    """
    c1=cos(t1)^(2/3)
    s1=sin(t1)^(2/3)
    c2=cos(t2)^(2/3)
    s2=sin(t2)^(2/3)
    return [[[c1,s1*c2,0],[s1*c2,s1*s2,0],[s1*s2,exp(-I*2*pi/3)*c1,0]],\
[[s1*s2,c1,exp(-I*2*pi/3)*s1*c2],[exp(I*2*pi/3)*c1,s1*c2,s1*s2],\
[s1*c2,s1*s2,c1]],[[0,s1*s2,c1],[0,c1,s1*c2],[0,exp(I*2*pi/3)*s1*c2,s1*s2]]]
\end{sageblock}\\
The use of the procedures for generating orthogonal hypermatrices
are illustrated bellow\\
\begin{sageblock}
theta = var('theta')
Q  = Orthogonal2x2x2Hypermatrix(theta)
Qt = HypermatrixCyclicPermute(Q)
Qtt= HypermatrixCyclicPermute(HypermatrixCyclicPermute(Q))
\end{sageblock}\\
Expressing $2\times2\times2$ orthogonal hypermatrices in term of
the free parameter $\theta$ we obtain 
\begin{equation}
\mathbf{Q}\left(\theta\right)=\sage{Q}
\end{equation}
\[
\circ\left(\mathbf{Q}\left(\theta\right),\,\left[\mathbf{Q}\left(\theta\right)\right]^{T^{2}},\,\left[\mathbf{Q}\left(\theta\right)\right]^{T}\right)=\mbox{HypermatrixProduct(Q, Qtt, Qt)}=
\]
\begin{equation}
\sage{HypermatrixProduct(Q,Qtt,Qt)}
\end{equation}
We also illustrate the output of the procedure implemented above for
generating parametrization for $3\times3\times3$ orthogonal hypermatrices\\
\begin{sageblock}
# Defining the Parametrization Variables
theta1,theta2=var('theta1,theta2')
c1=cos(theta1)^(2/3)
s1=sin(theta1)^(2/3)
c2=cos(theta2)^(2/3)
s2=sin(theta2)^(2/3)

# Parametrization of a orthogonal hypermatrix
U  = Orthogonal3x3x3Hypermatrix(theta1,theta2)
Ut = HypermatrixCyclicPermute(U)
Utt= HypermatrixCyclicPermute(HypermatrixCyclicPermute(U))
UUttUt = HypermatrixProduct(U,Utt,Ut)
for i in range(3):
    for j in range(3):
        for k in range(3):
            UUttUt[i][j][k] = (UUttUt[i][j][k]).simplify_exp()
\end{sageblock}\\
We verify that the obtained $3\times3\times3$ hypermatrix is indeed
orthogonal via the following computation
\begin{equation}
\left[\circ\left(\mathbf{U},\,\mathbf{U}^{T^{2}},\,\mathbf{U}^{T}\right)\right]_{0,0,0}=\sage{UUttUt[0][0][0]}
\end{equation}
\begin{equation}
\left[\circ\left(\mathbf{U},\,\mathbf{U}^{T^{2}},\,\mathbf{U}^{T}\right)\right]_{1,1,1}=\sage{UUttUt[1][1][1]}
\end{equation}
\begin{equation}
\left[\circ\left(\mathbf{U},\,\mathbf{U}^{T^{2}},\,\mathbf{U}^{T}\right)\right]_{2,2,2}=\sage{UUttUt[2][2][2]}
\end{equation}
\begin{equation}
\left[\circ\left(\mathbf{U},\,\mathbf{U}^{T^{2}},\,\mathbf{U}^{T}\right)\right]_{0,0,1}=\sage{UUttUt[0][0][1]}
\end{equation}
\begin{equation}
\left[\circ\left(\mathbf{U},\,\mathbf{U}^{T^{2}},\,\mathbf{U}^{T}\right)\right]_{0,0,2}=\sage{UUttUt[0][0][2]}
\end{equation}
\begin{equation}
\left[\circ\left(\mathbf{U},\,\mathbf{U}^{T^{2}},\,\mathbf{U}^{T}\right)\right]_{1,1,2}=\sage{UUttUt[1][1][2]}
\end{equation}
\begin{equation}
\left[\circ\left(\mathbf{U},\,\mathbf{U}^{T^{2}},\,\mathbf{U}^{T}\right)\right]_{1,1,0}=\sage{UUttUt[1][1][0]}
\end{equation}
\begin{equation}
\left[\circ\left(\mathbf{U},\,\mathbf{U}^{T^{2}},\,\mathbf{U}^{T}\right)\right]_{2,2,0}=\sage{UUttUt[2][2][0]}
\end{equation}
\begin{equation}
\left[\circ\left(\mathbf{U},\,\mathbf{U}^{T^{2}},\,\mathbf{U}^{T}\right)\right]_{2,2,1}=\sage{UUttUt[2][2][1]}
\end{equation}
\begin{equation}
\left[\circ\left(\mathbf{U},\,\mathbf{U}^{T^{2}},\,\mathbf{U}^{T}\right)\right]_{0,1,2}=\sage{UUttUt[0][1][2]}
\end{equation}
\begin{equation}
\left[\circ\left(\mathbf{U},\,\mathbf{U}^{T^{2}},\,\mathbf{U}^{T}\right)\right]_{1,0,2}=\sage{UUttUt[1][0][2]}
\end{equation}
In the remaining part of the package, we implement functions which
relates to genralizations to hypermatrices of the classic Cayley-Hamilton
theorem and to the notion of hypermatrix inversion. We first start
by implementing the function which creates a list of hypermatrices
corresponding to all the possible product composition of the input
hypermatrix $\mathbf{A}$.\\
\begin{sageblock}
def HypermatrixCayleyHamiltonList(A, n):
    """
    Outpts a list of hypermatrices of all product
    composition of order n from which it follows
    that n must be odd.

     EXAMPLES:
    ::
        sage: A = HypermatrixGenerate(2,2,2,'a')
        sage: L = HypermatrixCayleyHamiltonList(A,3)

    AUTHORS:
    - Edinah K. Gnang and Ori Parzanchevski
    """
    if n == 1:
        return [A]
    else:
        gu = []
        for i in range(1,n,2):
            for j in range(1,n-i,2):
                gu = gu + [HypermatrixProduct(g1,g2,g3) \
for g1 in HypermatrixCayleyHamiltonList(A,i) \
for g2 in HypermatrixCayleyHamiltonList(A,j) \
for g3 in HypermatrixCayleyHamiltonList(A,n-(i+j))]
        return gu
\end{sageblock}\\
It then becomes possible to establish that the dimension of the span
of hypermatrix composition powers is maximal for generic $2\times2\times2$
and $3\times3\times3$ hypermatrices. The lines of code bellow computes
hypermatrix compositions and stacks the resulting hypermatrices into
a square matrix and computing the determinant in order to assert that
the matrix is full rank.\\
\begin{sageblock}
# Initializing an orthogonal hypermatrix
A = Orthogonal2x2x2Hypermatrix(e/pi) 

# Initialization of the list
L = HypermatrixCayleyHamiltonList(A,1)+HypermatrixCayleyHamiltonList(A,3)+\
HypermatrixCayleyHamiltonList(A,5)+HypermatrixCayleyHamiltonList(A,7)

# Initializing the index variables
Indx = Set(range(len(L))).subsets(8)

# Initialization of the of the matrix
M = Matrix(RR,identity_matrix(8,8))
cnt = 0
for index in Indx:
    if cnt < 10:
        M = M*Matrix(RR,[HypermatrixVectorize(L[i]) for i in index])
        cnt= cnt+1
    else:
        break

# Defining the Parametrization Variables
c1=cos(e/pi)^(2/3)
s1=sin(e/pi)^(2/3)
c2=cos(pi/e)^(2/3)
s2=sin(pi/e)^(2/3)

# Defining the unitary hypermatrices
U=[[[c1,s1*c2,0],[s1*c2,s1*s2,0],[s1*s2, exp(-I*2*pi/3)*c1,0]],\
[[s1*s2,c1,exp(-I*2*pi/3)*s1*c2],\
[exp(I*2*pi/3)*c1,s1*c2,s1*s2],[s1*c2,s1*s2,c1]],\
[[0,s1*s2,c1],[0,c1,s1*c2],[0,exp(I*2*pi/3)*s1*c2,s1*s2]]]

Lu = HypermatrixCayleyHamiltonList(U,1)+HypermatrixCayleyHamiltonList(U,3)+\
HypermatrixCayleyHamiltonList(U,5)+HypermatrixCayleyHamiltonList(U,7)+\
HypermatrixCayleyHamiltonList(U,9)

# Initializing the index variables
Indxu = Set(range(len(Lu))).subsets(27)

# Initialization of the of the matrix
Mu = Matrix(CC,identity_matrix(27,27))
cntu = 0
for index in Indxu:
    if cntu < 5:
        Mu = Mu*Matrix(CC,[HypermatrixVectorize(Lu[i]) for i in index])
        cntu = cntu+1
    else:
        break
\end{sageblock}\\
The determinant of resulting matrix is given by 
\begin{equation}
\det\mathbf{M}=\sage{M.det()},
\end{equation}
furthermore for $3\times3\times3$ we have that 
\begin{equation}
\det\mathbf{M}^{\prime}=\sage{Mu.det()}
\end{equation}
The very last piece of the current package corresponds to the hypermatrix
pseudo-inversion procedure. The routine that we implement here will
be predominantly numerical. The notions of hypermatrix inverse pairs
was first proposed in the work of Battacharya and Mesner in \cite{BM2},
we follow up by implementing here numerical routine for the computation
of pseudo-inverse pairs for $2\times2\times2$ hypermatrices.

We first start by implementing a constraint formator procedure which
formats a list of linear constraints into a system of linear equation
in the canonical form $\mathbf{A}\cdot\mathbf{x}=\mathbf{b}$, the
constraint formator will be curcial for formating the linear constraints
which arise from the hypermatrix inversion constraints.\\
\begin{sageblock}
def ConstraintFormator(CnstrLst, VrbLst):
    """
    Takes as input a List of linear constraints 
    and a list of variables and outputs matrix 
    and the right hand side vector associate
    with the matrix formulation of the constraints.

    EXAMPLES:
    ::
        sage: x, y = var('x,y')
        sage: CnstrLst = [x+y==1, x-y==2]
        sage: VrbLst = [x, y]
        sage: [A,b] = ConstraintFormator(CnstrLst, VrbLst)

    AUTHORS:
    - Edinah K. Gnang and Ori Parzanchevski
    """
    # Initializing the Matrix
    A=Matrix(CC,len(CnstrLst),len(VrbLst),zero_matrix(len(CnstrLst),len(VrbLst)))
    b=vector(CC, [eq.rhs() for eq in CnstrLst]).column()
    for r in range(len(CnstrLst)):
        for c in range(len(VrbLst)):
            A[r,c] = diff((CnstrLst[r]).lhs(),VrbLst[c])
    return [A,b]

def ConstraintFormatorII(CnstrLst, VrbLst):
    """
    Takes as input a List of linear constraints 
    and a list of variables and outputs matrix 
    and the right hand side vector associate
    with the matrix formulation of the constraints.

    EXAMPLES:
    ::
        sage: x, y = var('x,y')
        sage: CnstrLst = [x+y==1, x-y==2]
        sage: VrbLst = [x, y]
        sage: [A,b] = ConstraintFormator(CnstrLst, VrbLst)

    AUTHORS:
    - Edinah K. Gnang and Ori Parzanchevski
    """
    # Initializing the Matrix
    A=Matrix(SR,len(CnstrLst),len(VrbLst),zero_matrix(len(CnstrLst),len(VrbLst)))
    b=vector(SR, [eq.rhs() for eq in CnstrLst]).column()
    for r in range(len(CnstrLst)):
        for c in range(len(VrbLst)):
            A[r,c] = diff((CnstrLst[r]).lhs(),VrbLst[c])
    return [A,b]
\end{sageblock}\\
Having implemented the constraint formator we are now ready to implement
the 3-hypermatrix pseudo-inversion subroutine. The following equation
constitutes the defining property of an inverse pairs for 3-hypermatrices
$\mathbf{A}$ and $\mathbf{B}$
\begin{equation}
\forall\:\mathbf{M}\in\mathbb{C}^{n\times n\times n},\quad\mathbf{M}=\circ\left(\circ\left(\mathbf{M},\mathbf{A},\mathbf{B}\right),\mathbf{U},\mathbf{V}\right)
\end{equation}
consequently as suggested by the equality above, the ordered pair
of hypermatrices $\left(\mathbf{U},\mathbf{V}\right)$ is said to
denote inverse pairs associated with the ordered hypermatrix pair
$\left(\mathbf{A},\mathbf{B}\right)$. in the case where no such hypermatrices
pairs exist for the pair $\left(\mathbf{A},\,\mathbf{B}\right)$ we
say that the pair $\left(\mathbf{A},\mathbf{B}\right)$ is non invertible
and in such case we may compute a pseudo-inverse inverse pair as follows.\\
\begin{sageblock}
def HypermatrixPseudoInversePairs(A,B):
    """
     Outputs the pseudo inverse pairs associated with the input pairs of matrices

    EXAMPLES:
    ::
        sage: A1=[[[0.1631135370902057,0.11600112072013125],[0.9823708115400902,0.39605960486710756]]\
,[[0.061860929755424676,0.2325542810173995],[0.39111210957450926,0.2019809359102137]]]
        sage: A2=[[[0.15508921433883183,0.17820377184410963],[0.48648171594508205,0.01568017636082064]]\
,[[0.8250247759993575,0.1938307874191597],[0.23867299119274843,0.3935578730402869]]]
        sage: [B1,B2]=HypermatrixPseudoInversePairs(A1,A2)

    AUTHORS:
    - Edinah K. Gnang and Ori Parzanchevski
    """
    sz = len(A)

    # Initializing the list of linear constraints
    CnstrLst = []

    # Initilizing the variable list
    Vrbls  = [var('ln_al'+str(i)+str(j)+str(k)) \
for i in range(sz) for j in range(sz) for k in range(sz)]+\
[var('ln_bt'+str(i)+str(j)+str(k)) for i in range(sz) for j in range(sz) \
for k in range(sz)]

    for m in range(sz):
        for p in range(sz):
            for n in range(sz):
                V=Matrix(CC, sz, sz, [(A[m][k1][k0])*(B[k0][k1][p]) \
for k0 in range(sz) for k1 in range(sz)]).inverse()
                CnstrLst=CnstrLst+[\
var('ln_al'+str(m)+str(n)+str(k1))+var('ln_bt'+str(k1)+str(n)+str(p))==\
ln(V[k1,n])  for k1 in range(sz)]
    [A,b]=ConstraintFormator(CnstrLst,Vrbls)

    # Importing the Numerical Python package
    # for computing the matrix pseudo inverse
    import numpy

    sln = matrix(numpy.linalg.pinv(A))*b
    R1 = HypermatrixGenerateAllZero(sz,sz,sz)
    for i in range(sz):
        for j in range(sz):
            for k in range(sz):
                R1[i][j][k] = exp(sln[i*sz^2+j*sz^1+k*sz^0,0])
    R2 = HypermatrixGenerateAllZero(sz, sz, sz)
    for i in range(sz):
        for j in range(sz):
            for k in range(sz):
                R2[i][j][k] = exp(sln[sz^3+i*sz^2+j*sz^1+k*sz^0,0])
    return [R1,R2]
\end{sageblock}\\
To illustrate how the procedure implemented above are used, we compute
for the inverse pair corresponding to the hypermatrix pair $\left(\mathbf{A}_{1},\,\mathbf{A}_{2}\right)$
specified bellow\\
\begin{sageblock}
# Building from the example mentioned in the implementation
# we consider the hypermatrices
A1=[[[0.1631135370902057,0.11600112072013125],\
[0.9823708115400902,0.39605960486710756]],\
[[0.061860929755424676,0.2325542810173995],\
[0.39111210957450926,0.2019809359102137]]]

A2=[[[0.15508921433883183,0.17820377184410963],\
[0.48648171594508205,0.01568017636082064]],\
[[0.8250247759993575,0.1938307874191597],\
[0.23867299119274843,0.3935578730402869]]]

# Numerical computation of the hypermatrix inverse pairs
[B1,B2]=HypermatrixPseudoInversePairs(A1,A2)

# To appreciate how good the numerical approximation of the
# inverse pair is we generate the generic symbolic hypermatrix M
M0 = HypermatrixGenerate(2,2,2,'m')

# We would want to compare the symbolic hypermatrix M to the product
M1 = HypermatrixProduct(M0,A1,A2)
M2 = HypermatrixProduct(M1,B1,B2)
\end{sageblock}\\
unfortunately the hypermatrix pair $\left(\mathbf{A}_{1},\,\mathbf{A}_{2}\right)$
(chosen here randomly above) admits no inverse pair and hence starting
from the generic symbolic hypermatrix $\mathbf{M}$ 
\begin{equation}
\mathbf{M}=\sage{M0}
\end{equation}
we illustrate the error induced by the pseudo-inversion by comparing
to $\mathbf{M}$ the 3-hypermatrix product computation $\circ\left(\circ\left(\mathbf{M},\mathbf{A}_{1},\mathbf{A}_{2}\right),\mathbf{B}_{1},\mathbf{B}_{2}\right)$
with entries given by 
\[
\left[\circ\left(\circ\left(\mathbf{M},\mathbf{A}_{1},\mathbf{A}_{2}\right),\mathbf{B}_{1},\mathbf{B}_{2}\right)\right]_{0,0,0}=\sage{((M2[0][0][0]).operands())[0]}+
\]
\begin{equation}
\sage{((M2[0][0][0]).operands())[1]}
\end{equation}
\[
\left[\circ\left(\circ\left(\mathbf{M},\mathbf{A}_{1},\mathbf{A}_{2}\right),\mathbf{B}_{1},\mathbf{B}_{2}\right)\right]_{0,0,1}=\sage{((M2[0][0][1]).operands())[0]}+
\]
\begin{equation}
\sage{((M2[0][0][1]).operands())[1]}
\end{equation}
\[
\left[\circ\left(\circ\left(\mathbf{M},\mathbf{A}_{1},\mathbf{A}_{2}\right),\mathbf{B}_{1},\mathbf{B}_{2}\right)\right]_{0,1,0}=\sage{((M2[0][1][0]).operands())[0]}+
\]
\begin{equation}
\sage{((M2[0][1][0]).operands())[1]}
\end{equation}
\[
\left[\circ\left(\circ\left(\mathbf{M},\mathbf{A}_{1},\mathbf{A}_{2}\right),\mathbf{B}_{1},\mathbf{B}_{2}\right)\right]_{0,1,1}=\sage{((M2[0][1][1]).operands())[0]}+
\]
\begin{equation}
\sage{((M2[0][1][1]).operands())[1]}
\end{equation}
\[
\left[\circ\left(\circ\left(\mathbf{M},\mathbf{A}_{1},\mathbf{A}_{2}\right),\mathbf{B}_{1},\mathbf{B}_{2}\right)\right]_{1,0,0}=\sage{((M2[1][0][0]).operands())[0]}+
\]
\begin{equation}
\sage{((M2[1][0][0]).operands())[1]}
\end{equation}
\[
\left[\circ\left(\circ\left(\mathbf{M},\mathbf{A}_{1},\mathbf{A}_{2}\right),\mathbf{B}_{1},\mathbf{B}_{2}\right)\right]_{1,0,1}=\sage{((M2[1][0][1]).operands())[0]}+
\]
\begin{equation}
\sage{((M2[1][0][1]).operands())[0]}
\end{equation}
\[
\left[\circ\left(\circ\left(\mathbf{M},\mathbf{A}_{1},\mathbf{A}_{2}\right),\mathbf{B}_{1},\mathbf{B}_{2}\right)\right]_{1,1,0}=\sage{((M2[1][1][0]).operands())[0]}+
\]
\begin{equation}
\sage{((M2[1][1][0]).operands())[1]}
\end{equation}
\[
\left[\circ\left(\circ\left(\mathbf{M},\mathbf{A}_{1},\mathbf{A}_{2}\right),\mathbf{B}_{1},\mathbf{B}_{2}\right)\right]_{1,1,1}=\sage{((M2[1][1][1]).operands())[0]}+
\]
\begin{equation}
\sage{((M2[1][1][1]).operands())[1]}
\end{equation}

\subsection{Hypermatrix class}

As a summary for the hypermatrix package for the convenience of the
user we encapsulate all the pieces into a single class all the precedures
implemented above. This is particularly useful for the purpose of
setting up computer experiments with the proposed hypermatrix package.
\\
\begin{sageblock}
class HM:
    """HM class"""
    def __init__(self,*args):
		# Single argument class constructor specification.
        if len(args) == 1:
            inp = args[0]
            if type(inp)==type(Matrix(SR,2,1,[var('x'),var('y')])) or \
type(inp)==type(Matrix(RR,2,1,[1,2])) or type(inp)==type(Matrix(CC,2,1,[1,1])):
                self.hm=DiagonalHypermatrix(inp)
            elif type(inp) == list:
                self.hm = inp
            else:
                raise ValueError, \
"Expected either a list or and an object of type Matrix"
            return
        # Two or more arguments class constructor
        s = args[-1]
        dims = args[:-1]
        if s == 'one':
            self.hm = apply(HypermatrixGenerateAllOne, dims)
        elif s == 'zero':
            self.hm = apply(HypermatrixGenerateAllZero, dims)
        elif s == 'ortho':
            if len(dims) == 1:
                self.hm=Orthogonal2x2x2Hypermatrix(dims[0])
            elif len(dims) == 2:
                self.hm=Orthogonal3x3x3Hypermatrix(dims[0],dims[1])
            else:
                raise ValueError,\
"ortho not supported for order %d tensors" % len(dims)
        elif s == 'perm':
            self.hm=HypermatrixPermutation(dims[0])
        elif s == 'kronecker':
            self.hm=HypermatrixKroneckerDelta(dims[0])
        elif s == 'sym':
            if len(dims) == 2:
                self.hm=SymHypermatrixGenerate(dims[0],dims[1])
            else :
                raise ValueError,\
"kronecker not supported for order %d tensors" % len(dims)
        else:
            self.hm=apply(HypermatrixGenerate, args)

    def __repr__(self):
        return `self.hm`

    def __add__(self, other):
        return GeneralHypermatrixAdd(self,other)

    def __neg__(self):
        return GeneralHypermatrixScale(self.hm,-1)

    def __sub__(self, other):
        return GeneralHypermatrixAdd(self, GeneralHypermatrixScale(other,-1))

    def __mul__(self, other):
        if other.__class__.__name__=='HM':
            return HM(GeneralHypermatrixHadamardProduct(self,other))
        elif other.__class__.__name__=='tuple':
            # This function takes a a list as intput
            l = other
            return GeneralHypermatrixProduct(self,*l)
        else:
            return GeneralHypermatrixScale(self,other)

    def __rmul__(self, a):
        return self*a

    def __getitem__(self,i):
        if i.__class__.__name__=='tuple':
            tmp = self.hm
            for j in i:
                tmp = tmp[j]
            return tmp

    def __setitem__(self, i, v):
        if   i.__class__.__name__=='tuple':
            tmp = self.hm
            while len(i)>1:
                tmp = tmp[i[0]]
                i = i[1:]
            tmp[i[0]] = v

    def __call__(self, *inpts):
        # This function takes a a list as intput
        return GeneralHypermatrixProduct(self, *inpts)

    def hprod(self,*inpts):
        # This function takes a a list as intput
        return GeneralHypermatrixProduct(self,*inpts)

    def hprod3b(self, b, c, t):
        return HM(HypermatrixProductB(self.hm, b.hm, c.hm, t.hm))

    def elementwise_product(self,B):
        return GeneralHypermatrixHadamardProduct(self,B)

    def elementwise_exponent(self,s):
        return GeneralHypermatrixExponent(self,s)

    def elementwise_base_exponent(self,s):
        return GeneralHypermatrixBaseExponent(self,s)

    def transpose(self, i=1):
        t = Integer(mod(i, self.order()))
        A = self
        for i in range(t):
            A = GeneralHypermatrixCyclicPermute(A)
        return A

    def nrows(self):
        return len(self.hm)

    def ncols(self):
        return len(self.hm[0])

    def ndpts(self):
        return len(self.hm[0][0])

    def n(self,i):
        tmp = self.listHM()
        for j in range(i):
            tmp = tmp[0]
        return len(tmp)

    def list(self):
        lst = []
        l = [self.n(i) for i in range(self.order())]
        # Main loop canonicaly listing the elements
        for i in range(prod(l)):
            entry = [mod(i,l[0])]
            sm = Integer(mod(i,l[0]))
            for k in range(len(l)-1):
                entry.append(Integer(mod(Integer((i-sm)/prod(l[0:k+1])),l[k+1])))
                sm = sm+prod(l[0:k+1])*entry[len(entry)-1]
            lst.append(self[tuple(entry)])
        return lst

    def listHM(self):
        return self.hm

    def cayley_hamilton_list(self,n):
        tmp = HypermatrixCayleyHamiltonList(self.hm,n)
        return [HM(h) for h in tmp]

    def cayley_hamilton_mtrx(self,itr,bnd):
        tmp = []
        for i in range(itr):
            tmp = tmp + HypermatrixCayleyHamiltonList(self.hm, 2*i+1)
        return Matrix([HM(h).list() for h in tmp[0:bnd]])

    def order(self):
        cnt = 0
        H = self.listHM()
        while type(H) == type([]):
            H = H[0]
            cnt = cnt+1
        return cnt
\end{sageblock}\\
We implement also some additional auxiliary special functions specifically
used by the class for dealing with hypermatrices of order greater
then $3$. We start by implementing a general hypermatrix product
operation which incorporate matrix and 3-hypermatrix products as special
cases and therfore captures the full Bhattacharya-Mesner algebra.\\
\begin{sageblock}
def GeneralHypermatrixProduct(*args):
    # Initialization of the list specifying the dimensions of the output
    l = [(args[i]).n(i) for i in range(len(args))]
    # Initializing the input for generating a symbolic hypermatrix
    inpts = l+['zero']
    # Initialization of the hypermatrix
    Rh = HM(*inpts)
    # Main loop performing the assignement
    for i in range(\
prod([(args[j]).n(Integer(mod(j+1,len(args)))) for j in range(len(args))])):
        entry = [mod(i,l[0])]
        sm = Integer(mod(i,l[0]))
        for k in range(len(l)-1):
            entry.append(Integer(mod(Integer((i-sm)/prod(l[0:k+1])),l[k+1])))
            sm = sm+prod(l[0:k+1])*entry[len(entry)-1]
        if len(args)<2:
            raise ValueError, "The number of operands must be >= 2"
        elif len(args) >= 2:
            Rh[tuple(entry)]=sum(\
[prod([args[s][tuple(entry[0:Integer(mod(s+1,len(args)))]+\
[t]+entry[Integer(mod(s+2,len(args))):])] for s in range(len(args)-2)]+\
[args[len(args)-2][tuple(entry[0:len(args)-1]+[t])]]+\
[args[len(args)-1][tuple([t]+entry[1:])]]) for t in range((args[0]).n(1))])
    return Rh
\end{sageblock}\\
We also implement the more generally cyclic action on arbitrary order
hypermatrices.\\
\begin{sageblock}
def GeneralHypermatrixCyclicPermute(A):
    # Initialization of the list specifying the dimensions of the output
    l = [A.n(i) for i in range(A.order())]
    l = l[1:]+[l[0]]
    # Initializing the input for generating a symbolic hypermatrix
    inpts = l+['r']
    # Initialization of the hypermatrix
    Rh = HM(*inpts)
    # Main loop performing the transposition of the entries
    for i in range(prod(l)):
        entry = [mod(i,l[0])]
        sm = Integer(mod(i,l[0]))
        for k in range(len(l)-1):
            entry.append(Integer(mod(Integer((i-sm)/prod(l[0:k+1])),l[k+1])))
            sm = sm+prod(l[0:k+1])*entry[len(entry)-1]
        # Performing the transpose
        Rh[tuple(entry)]=A[tuple([entry[len(entry)-1]]+entry[:len(entry)-1])]
    return Rh
\end{sageblock}\\
similarly the scaling function will be given by\\
\begin{sageblock}
def GeneralHypermatrixScale(A,s):
    # Initialization of the list specifying the dimensions of the output
    l = [A.n(i) for i in range(A.order())]
    # Initializing the input for generating a symbolic hypermatrix
    inpts = l+['r']
    # Initialization of the hypermatrix
    Rh = HM(*inpts)
    # Main loop performing the transposition of the entries
    for i in range(prod(l)):
        entry = [mod(i,l[0])]
        sm = Integer(mod(i,l[0]))
        for k in range(len(l)-1):
            entry.append(Integer(mod(Integer((i-sm)/prod(l[0:k+1])),l[k+1])))
            sm = sm+prod(l[0:k+1])*entry[len(entry)-1]
        # Performing the computation
        Rh[tuple(entry)]=s*A[tuple(entry)]
    return Rh
\end{sageblock}\\
the function which compute the exponentiation of the entries for arbitrary
order hypermatrices is given bellow\\
\begin{sageblock}
def GeneralHypermatrixExponent(A,s):
    # Initialization of the list specifying the dimensions of the output
    l = [A.n(i) for i in range(A.order())]
    # Initializing the input for generating a symbolic hypermatrix
    inpts = l+['r']
    # Initialization of the hypermatrix
    Rh = HM(*inpts)
    # Main loop performing the transposition of the entries
    for i in range(prod(l)):
        entry = [mod(i,l[0])]
        sm = Integer(mod(i,l[0]))
        for k in range(len(l)-1):
            entry.append(Integer(mod(Integer((i-sm)/prod(l[0:k+1])),l[k+1])))
            sm = sm+prod(l[0:k+1])*entry[len(entry)-1]
        # Performing computation
        Rh[tuple(entry)]=(A[tuple(entry)])^s
    return Rh
\end{sageblock}\\
for accounting for the non commutativeity of the exponentiation operation
we implement the other function\\
\begin{sageblock}
def GeneralHypermatrixBaseExponent(A,s):
# Initialization of the list specifying the dimensions of the output
    l = [A.n(i) for i in range(A.order())]
    # Initializing the input for generating a symbolic hypermatrix
    inpts = l+['r']
    # Initialization of the hypermatrix
    Rh = HM(*inpts)
    # Main loop performing the transposition of the entries
    for i in range(prod(l)):
        entry = [mod(i,l[0])]
        sm = Integer(mod(i,l[0]))
        for k in range(len(l)-1):
            entry.append(Integer(mod(Integer((i-sm)/prod(l[0:k+1])),l[k+1])))
            sm = sm+prod(l[0:k+1])*entry[len(entry)-1]
        # Performing computation
        Rh[tuple(entry)]=s^(A[tuple(entry)])
    return Rh
\end{sageblock}\\
Similarly arbitrary order hypermatrix addition is implemented bellow
as\\

\begin{sageblock}
def GeneralHypermatrixAdd(A,B):
    # Initialization of the list specifying the dimensions of the output
    l = [A.n(i) for i in range(A.order())]
    s = [B.n(i) for i in range(B.order())]
    # Testing the dimensions
    x = var('x')
    if(sum([l[i]*x^i for i in range(len(l))])==sum(\
[s[i]*x^i for i in range(len(s))])):
        # Initializing the input for generating a symbolic hypermatrix
        inpts = l+['r']
        # Initialization of the hypermatrix
        Rh = HM(*inpts)
        # Main loop performing the transposition of the entries
        for i in range(prod(l)):
            entry = [mod(i,l[0])]
            sm = Integer(mod(i,l[0]))
            for k in range(len(l)-1):
                entry.append(Integer(mod(Integer((i-sm)/prod(l[0:k+1])),l[k+1])))
                sm = sm+prod(l[0:k+1])*entry[len(entry)-1]
            Rh[tuple(entry)]=A[tuple(entry)]+B[tuple(entry)]
        return Rh
    else:
        raise ValueError,\
"The Dimensions of the input hypermatrices must match."
\end{sageblock}\\
quite similarly the Hadamard product is given by\\
\begin{sageblock}
def GeneralHypermatrixHadamardProduct(A,B):
# Initialization of the list specifying the dimensions of the output
    l = [A.n(i) for i in range(A.order())]
    s = [B.n(i) for i in range(B.order())]
    # Testing the dimensions
    x = var('x')
    if(sum([l[i]*x^i for i in range(len(l))])==sum(\
[s[i]*x^i for i in range(len(s))])):
        # Initializing the input for generating a symbolic hypermatrix
        inpts = l+['r']
        # Initialization of the hypermatrix
        Rh = HM(*inpts)
        # Main loop performing the transposition of the entries
        for i in range(prod(l)):
            entry = [mod(i,l[0])]
            sm = Integer(mod(i,l[0]))
            for k in range(len(l)-1):
                entry.append(Integer(mod(Integer((i-sm)/prod(l[0:k+1])),l[k+1])))
                sm = sm+prod(l[0:k+1])*entry[len(entry)-1]
            Rh[tuple(entry)]=A[tuple(entry)]+B[tuple(entry)]
        return Rh
    else:
        raise ValueError,\
"The Dimensions of the input hypermatrices must match."
\end{sageblock}\\
As a way of illustrating the existence of arbitrary order hypermatrices
we implement a procedure for parametrizing ( albeit somewhat redundantly
in the number of variables ) hypermatrices of the size $2\times2\times2\times\cdots\times2$.
and therefore providing a constructif proof of existence of arbitrary
order orthogonal hypermatrices.\\
\begin{sageblock}
def GeneralOrthogonalHypermatrix(od):
    # Initializing the hypermatrix
    Q = apply(HM,[2 for i in range(od)]+['q'])

    # Initilizing the list of variable
    VrbLst = Q.list()

    # Reinitializing of Q by exponentiation
    Q = Q.elementwise_base_exponent(e)

    # Computing the product
    Eq = apply(GeneralHypermatrixProduct,[Q.transpose(j) for j in range(od,0,-1)])

    # Writting up the constraints
    LeQ = (Set(Eq.list())).list()

    # Removing the normalization constraints
    LeQ.remove(e^(od*var('q'+''.join(['0' for i in range(od)])))+\
e^(od*var('q01'+''.join(['0' for i in range(od-2)]))))
    LeQ.remove( e^(od*var('q10'+''.join(['1' for i in range(od-2)])))+\
e^(od*var('q'+''.join(['1' for i in range(od)]))))

    # Filling up the linear constraints
    CnstrLst= []
    for f in LeQ:
        CnstrLst.append(\
ln((f.operands())[0]).simplify_exp()-I*pi-ln((f.operands())[1]).simplify_exp()==0)

    # Directly solving the constraints
    Sl = solve(CnstrLst,VrbLst)

    # Main loop performing the substitution of the entries
    Lr = [var('r'+str(i)) for i in range(1,2^od+1)]
    l = [Q.n(i) for i in range(Q.order())]
    for i in range(prod(l)):
        # Turning the index i into an hypermatrix array location
        # using the decimal encoding trick
        entry = [mod(i,l[0])]
        sm = Integer(mod(i,l[0]))
        for k in range(len(l)-1):
            entry.append(Integer(mod(Integer((i-sm)/prod(l[0:k+1])),l[k+1])))
            sm = sm+prod(l[0:k+1])*entry[len(entry)-1]
        Q[tuple(entry)]=Q[tuple(entry)].subs(\
dict(map(lambda eq: (eq.lhs(),eq.rhs()), Sl[0]))).simplify_exp()
    return Q
\end{sageblock}

\section*{Acknowledgments}

We would like to thank the IAS for providing excellent working conditions.
This material is based upon work supported by the National Science
Foundation under agreements Princeton University Prime Award No. CCF-0832797
and Sub-contract No. 00001583. Any opinions, findings and conclusions
or recommendations expressed in this material are those of the authors
and do not necessarily reflect the views of the National Science Foundation.

\end{document}